\documentclass{jpp}
\usepackage{graphicx}

\usepackage[utf8]{inputenc}
\usepackage[T1]{fontenc}
\usepackage{amsmath}

\shorttitle{Boundary Layers, Noise, and Line-tying}
\shortauthor{A.~B. Hassam and Y.-M. Huang}

\title{Boundary layers and noise in magnetized plasmas line-tied at conducting surfaces}

\author{A.~B. Hassam\aff{1}
  \corresp{\email{hassam@umd.edu}}
 \and Yi-Min Huang\aff{2}}

\affiliation{\aff{1} Department of Physics, University of Maryland, College Park, Maryland, 20740, USA
\aff{2}Department of Astrophysical Sciences, Princeton University, Princeton, New Jersey, 08544, USA}

\begin{document}

\maketitle

\begin{abstract}
In magnetized plasma situations where magnetic fields intersect massive conducting boundaries, "line-tied" boundary conditions are often used, analytically and in numerical simulations. For ideal MHD plasmas, these conditions are arrived at given the relatively long time scales for magnetic fields penetrating resistively into good conductors.  Under line-tied boundary conditions, numerical simulations often exhibit what could be construed as numerical "noise" emanating from the boundaries.  We show here that this "noise" is real.  By combining numerical and analytical methods, we highlight the existence of boundary layers near the conductors and confirm the appearance of short wavelength structures riding on long wavelength features. We conclude that for numerical fidelity the boundary layers need to be resolved.  Boundary layer widths scale as the square root of the plasma $\beta$.    
\end{abstract}

\section{Introduction}
There are many situations where strong magnetic fields in plasma intersect massive, inert, conducting boundaries.   One example is emerging magnetic flux from the solar surface;  here the solar surface and below is massive as well as conducting \citep{Parker1972,HoodP1979,Antiochos1987,LionelloVEM1998,Priest2014,HuangZS2006}.  As another example are  laboratory magnetized plasma experiments where magnetic lines meet vacuum vessel walls. 
In both these situations, the plasma MHD timescales are much shorter than resistive penetration times into the conductor.  Thus, the field lines are effectively line-tied.

In numerical solutions of magnetized plasma for such situations, so-called “line-tied” boundary conditions are implemented by assuming that both the tangential electric field and the perturbed normal magnetic field be zero at the boundary.  For MHD, since the flows are dominantly $E \times B$, this translates to zero tangential flows (where strong fields cut into the boundary).  In addition, if the boundary is assumed effectively impervious, the normal flow is also set to zero, and the normal derivative of the density is zero (“hard-wall” boundaries).   Finally, the normal derivative of the tangential perturbed magnetic field is assumed to be zero.   [\citep{Hood1} has considered line-tied boundary conditions in detail.  He distinguishes between impervious boundaries and boundaries that allow normal flow through.  In this paper, we assume Hood's impervious choice.]

\begin{figure}
  \centering
  \includegraphics[width=0.6\textwidth]{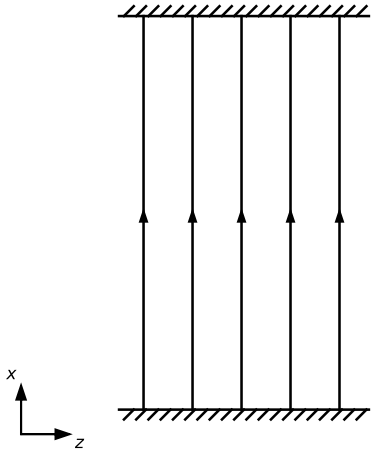}
  \caption{Uniform $\boldsymbol B = B_{0x} \boldsymbol{\hat x}$ magnetic field in between conducting plates.  All tests done in this paper refer to linear perturbations about this MHD system.  Fast and slow MHD modes are expected.  Domain of $x=[-1,1]$, $z$ assumed periodic.}
\label{fig:line-tied}
\end{figure}

As an example, consider a 1D shear Alfv\'en wave set up between two conducting plates at $x=-1$ and $x=+1$,  as shown in Fig.~\ref{fig:line-tied}.  There is an equilibrium uniform magnetic field, $\boldsymbol{B}_0=B_{0x}\boldsymbol{\hat{x}}$.  
An initial condition on the $z$-directed plasma flow, $u_z = \sin(Kx)$, $K=\pi/2$, is given.  In a code we describe later \citep{GuzdarDMHL1993}, which uses a fourth order finite difference scheme, two “ghost points” outside of the domain need to be specified.  The simplest prescription, consistent with Line-tied/Hard-wall (LTHW) boundary conditions, is to symmetrize $B_z$ and antisymmetrize $u_z$ across the boundaries.  Further, $u_z(x=\pm 1)$ are set to zero, while $B_z(x= \pm 1)$ are stepped.  Corresponding boundary conditions for other possibly involved variables would be antisymmetric $u_x$ and $B_x$, and symmetric density, $n$.  The 1D Alfv\'en wave, in a numerical solution as initialized above, works very well, and, in particular, is noise-free.  

The situation, however, is more complex in the case of a 2D or 3D situation. A system set up such as in Fig.~\ref{fig:line-tied}, if subjected to 2D or 3D perturbations, can be shown to generate small scale structures.  While the structures may remain small amplitude, they sometimes devolve into apparent numerical instability.  The latter consequence is especially true for nonlinear problems.  It is well recognized by workers in the field that this type of "noise" is often observed.  A remedy sometimes suggested is to place a strong but narrow diffusion layer very close to the boundary.  While this is apparently sometimes workable, there is a risk of slippage in line-tying (to the extent that line-tying is an important part of the matter at hand).

In this paper, we explore a system, such as in Fig.~\ref{fig:line-tied}, both analytically and numerically, to understand the origin of the "noise".  
In particular, we study the linear ideal MHD normal modes of the system of Fig.~\ref{fig:line-tied}, with LTHW boundary conditions.  From analytic solutions, we find two characteristics of the normal modes:

(1)  the eigenfunctions show the presence of sharp boundary layers near the conducting walls;

(2)  the eigenfunctions show a two-space scale structure, specifically a high wave number "ripple" riding on low-wavenumber envelope solutions.

This kind of multiscale structure could easily be misconstrued as "noise".  In fact, if the boundary layers are not sufficiently numerically resolved, this could result in real numerical noise.  Our conclusion from these findings is that high resolution numerical analysis may be generally necessary for these types of problems.  

In the next four sections, we analytically calculate the eigenmodes of the system.  This calculation, we will show, requires boundary layer and multiscale methods.  In Sec. 6, we run several numerical simulations to bolster and confirm the analytic solutions. We compare different numerical solution methods to quantify the grid resolution necessary to address line-tied problems.





\section{Equations}\label{sec:equations}
As mentioned, we investigate the system shown in Fig.~\ref{fig:line-tied}.   Ideal MHD equations in 2D are considered.  In usual notation, the governing equations are: 
\begin{equation}
    \partial_t n + \nabla \cdot n\boldsymbol{u} = 0  \label{density} 
\end{equation}
\begin{equation}
    nM d_t \boldsymbol{u}  = - T \nabla n  +  j_y\boldsymbol{\hat{y}} \times \boldsymbol{B} 
    \label{momentum}
\end{equation}
\begin{equation}
    j_y = \boldsymbol{\hat{y}} \cdot \nabla \times \boldsymbol{B} 
    \label{jy}
\end{equation}
\begin{equation}
    \partial_t \boldsymbol{B}  = \nabla \times (\boldsymbol{u} \times \boldsymbol{B})
    \label{induction}    
\end{equation}
\begin{equation}
    \nabla \cdot \boldsymbol{B} = 0
    \label{divB}    
\end{equation}
where the temperature, $T$, is assumed isothermal. The vector $\boldsymbol{B}$ is assumed to be in the $x-z$ plane. (For the purposes of this paper, the system in Fig.~\ref{fig:line-tied} will be normalized as follows: we set the distance between the plates to 2, where the domain of $x$ is [-1,1].  Also, we set the Alfv\'en speed to unity.  In particular, $B_{0x}=1$, and $n_0M=1$. Thus energies are normalized to the magnetic energy, and, in normalized units, $T$ is the plasma $\beta$.)    While these equations are nonlinear, and what we present in this paper pertains also to nonlinear conditions, it suffices to demonstrate our findings in the linear regime.  Accordingly, from Eqns (\ref{density})--(\ref{divB}), we obtain a linearized system in the variables $n$, $u_x$, $u_z$, and $B_x$, given by 
\begin{equation}
    \partial_t n  +  \partial_x u_x  =  - i k u_z 
    \label{linear_n}
\end{equation}
\begin{equation}
    \partial_t u_x  =  - T \partial_x n  
    \label{linear_ux}
\end{equation}
\begin{equation}
    \partial_t u_z  =  -i k T n  +  (i/k) \nabla^2 B_x
    \label{linear_uz}
\end{equation}
\begin{equation}
    \partial_t B_x  = - i k u_z,
    \label{linear_bx}
\end{equation}
where $\partial_x B_x  =  -i k B_z$. The Laplacian $\nabla^2$ is defined as $\partial_x^2 - k^2$.  Here, we have assumed solutions of the form $\exp(ikz)$, where $k = 2\pi/L$.  The quantities $n$, $u_x$, $u_y$, $u_z$, $B_x$, $B_y$, $B_z$, are all small perturbations, compared to $n_0$ and $B_{0x}$.   These equations are fourth order in $d/dt$ and thus yield four ideal MHD eigenmodes, two fast modes and two slow modes.  Generally speaking, with the LTHW boundary conditions as above, this system will exhibit the type of noise discussed above.

\section{Mismatch}\label{sec:mismatch}

We begin by noting that if $k$ is set to zero in Eqs (\ref{linear_n})--(\ref{linear_bx}), the system is 1-D.  In that case, the modes clearly separate into the sound wave and the Alfven wave.  In particular, the first two equations can yield a standing sound wave,  parallel to the B field, coupling $n$ and  the parallel flow $u_x$. The boundary conditions are "purely" hard-wall: in particular, a sinusoidal, i.e., $\sin(Kx)$, density perturbation, (given the symmetric boundary conditions at $x=-1$ and $x=+1$) “forces” a co-sinusoidal behavior in $u_x$, i.e., $\cos(Kx)$.  The latter is consistent with the hard-wall antisymmetric boundary conditions for $u_x$.  
Similarly, the last two equations set up an Alfven wave coupling $uz$ to $B_x$.   Both these variables are co-sinusoidal, ie $\cos(Kx)$, consistent with the antisymmetric line-tied boundary conditions for each.

If, however, $k$ is non-zero, the eigenfunctions are more intricate.  We will show this in the next two sections.  Here, we first point out that there is a “mismatch”, between slow and fast modes, in their spatial sinusoidal parities.  Consider the slow wave.  This is a  sound like mode, propagating parallel to the B field.  The sinusoidal parities were described above.  However, if one examines Eq (\ref{linear_n}), within the assumption of the parities of $n$ and $u_x$ just mentioned, the parity that would “fit” the variable $u_z$, from Eq (\ref{linear_n}), would be sinusoidal, ie, $\sin(Kx)$.  But this parity violates the assumption of antisymmetric boundary conditions for $u_z$.  This violation is what we refer to as a “mismatch” between the slow and fast modes.  We will show that this type of mismatch gives rise to a boundary layer in $u_z$, since $u_z=0$ is enforced at the plates.  


A similar mismatch also arises in the fast mode for $k$ nonzero.  The fast mode is generally an Alfv\'enic mode, involving largely $B_x$ (or $B_z$) and $u_z$.  As already discussed, this has $u_z\sim \cos(Kx)$, being antisymmetric at the plates; while $B_x \sim \cos(Kx)$, consistent with $B_x$ also being antisymmetric at the plates.  However, from Eq (\ref{linear_uz}), the parities as deduced for $u_z$ and $B_x$ would force $n$ to be $\cos(Kx)$, which is antisymmetric at the plates, and so violates the symmetric boundary condition requirement for $n$ at the plates.  

We will now show below that these mismatches play a role in boundary layer formation in these modes, leading to short scale structure.

\section{Analytical Solution --- slow mode}

We proceed to obtain analytically the slow and fast eigenmodes for the system in Fig.~\ref{fig:line-tied}.  We begin by rewriting the four Eqs (\ref{linear_n})--(\ref{linear_bx}) as two coupled equations.  We use (\ref{linear_ux}) in (\ref{linear_n}) and (\ref{linear_bx}) in (\ref{linear_uz}) to get the following 2-field equations for $n$ and $B_x$

\begin{equation}
    (\gamma^2 - T\partial_x^2)n = \gamma^2 B_x
    \label{2n}
\end{equation}
\begin{equation}
    (\gamma^2 - \nabla^2)B_x = - k^2Tn
    \label{2bx}
\end{equation}
where solutions of the form $\exp(\gamma t) \exp(ikz)$ are assumed, and $\gamma$ is the eigenvalue.

We now make the low-beta assumption $T \ll 1$.  We obtain a reduced description for the slow mode in the $T \ll 1$ regime.  Such reduction has the advantage of practically decoupling the slow and fast modes, allowing for easier equations that need to be solved.   (Our conclusions in this paper, we will show, are independent of this reduction.)    Note, the limit $T \ll 1$ ensures that the sound speed is subdominant to the Alfv\'en speed. 

  We begin with the slow mode, wherein we anticipate that $\gamma^2 \ll 1$ (the slow mode is sub-Alfv\'enic).  In this case, from Eq (\ref{2bx}), we can discard the $\gamma^2$ term; but we order the Laplacian to be $O(1)$, i.e., long wavelengths.  Since the RHS is proportional to $T \ll 1$, we conclude that $B_x$ must be small to allow an optimal balance in that equation.  Thus, for $T \ll 1$, we have from Eq (\ref{2bx})
\begin{equation}
    \nabla^2 B_x = k^2 T n.
\label{slo1}
\end{equation}
From Eq (\ref{2n}), comparing the two $\gamma^2$ terms, we see that the term on the RHS must be neglected since $B_x$ is small, making the RHS quadratically small.  The remaining two terms on the LHS are of the same order, and so Eq (\ref{2n}) reduces as 
\begin{equation}
    (\gamma^2 - T\partial_x^2)n = 0.     \label{slo2}
\end{equation}

Eq (\ref{slo2}) is now an eigenvalue equation for $n(x)$ with eigenvalue $\gamma^2$.  The $n$ eigenfunction and the eigenvalue, invoking the appropriate hard-wall boundary condition, are 
\begin{equation}
    n = \sin(Kx),\text{    }\gamma^2 = - K^2 T.
\label{slo3}
\end{equation}
This is a long wavelength sound wave.  

Turning to Eq (\ref{slo1}), this is now an inhomogeneous equation for $B_x$.  The solution, constructed from a homogeneous solution and an added particular solution, and satisfying the appropriate line-tied boundary condition, is
\begin{equation}
    B_x = - T [ \sin(Kx) - \sinh(kx)/\sinh(k)].
\label{slo4}
\end{equation}

\begin{figure}
  \centering
  \includegraphics[width=0.6\textwidth]{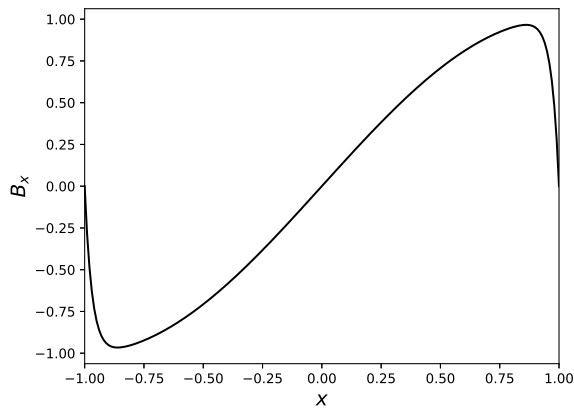}
  \caption{$B_x(x)$ plotted from Eq (\ref{slo4}).  $k/K$ chosen to be 20 to highlight boundary layers at the $x=\pm 1$ plates.}
\label{fig:slowmode}
\end{figure}
Note in the solution the appearance of hyperbolic behaviour, which brings into doubt the use of simple symmetric/antisymmetric boundary conditions.  In fact, if we assume $k \gg K$, the appearance of a boundary layer is clearly evident, as shown in Fig.~\ref{fig:slowmode}. The Figure shows $B_x(x)$ plotted from Eq.~(\ref{slo4}).  $k/K$ is chosen to be 20 to highlight the boundary layer at the $x=\pm1$ plates. Note that for $k/K \gg 1$, $B_x$ approaches the boundaries at $x=\pm1$ as symmetric but then there is a boundary layer near the plate enforced by the zero boundary condition at $x=\pm1$.  As is evident, 
the "mismatch" discussed earlier results in the boundary layer.  

We note here that we have not used any boundary layer analysis to arrive at the solution for $B_x$ (Eq. $\ref{slo4}$).   However, we clearly see a boundary layer in Fig.~\ref{fig:slowmode}, provided $k/K \gg 1$.  This is because for our calculation we assumed an "optimal ordering", $k \sim K$, and we were fortunate enough to be able to solve the equation.  Alternatively, if we had assumed $k/K$ large, right from the start, we would have found the $\sin(Kx)$ solution but not the $\sinh(kx)$ part.  We would have then had a "mismatched" boundary condition.  This would have precipitated a boundary layer analysis, leading to an "inner
solution" matching asymptotically with the "outer solution" \citep{Bender}.   We have indeed verified this behavior.  We do not show this calculation here, except to mention it, as we already have a uniformly valid solution everywhere. In the next section (the fast mode) we will indeed be forced to resort to boundary layer methods.

It is worth noting here that the emergence of a boundary layer in the $k/K$ large case has an important consequence.   In particular, the "reduced equations" of MHD \citep{Strauss1976} are derived under the assumption of long parallel wavelengths compared with perpendicular wavelengths, generally stated as $k_{\parallel} \ll k_{\perp}$.  But this is exactly our condition above.  Thus, the application of the reduced equations at conducting surfaces should require boundary layer analysis near the conductors.  This is a situation indeed encountered.  The appearance and necessity of a boundary layer for reduced equations has indeed been reported previously \citep{ScheperH1999}.  In the latter paper, it was shown that low frequency motions near the solar conducting surface would require a boundary layer analysis (thus necessitating the use of non-reduced equations in the layer.) 

As a final comment on the slow mode, we note that an equilibrium solution ($\partial_t = 0$), with density having $k$ and $K$ variations, can be readily found.  This too has boundary layers at the plates. 

\section{Analytical Solution --- fast mode}

We now turn to the fast mode.  We begin as in Sec 4 with Eqs (\ref{2n}) and (\ref{2bx}).  This time, we order $\gamma^2$ as $O(1)$, as we anticipate an Alfv\'en mode.  Since $T \ll 1$, and we assume $\partial_{x}^2$ is $O(1)$ for long wavelengths, Eq (\ref{2n}) reduces to
\begin{equation}
    \gamma^2 n = \gamma^2 B_x  .
    \label{f1}
\end{equation}
This yields $n = B_x$.  Using the latter in Eq (\ref{2bx}), the RHS of (\ref{2bx}) must be discarded, being small, leading to 

\begin{equation}
    (\gamma^2 - \nabla^2)B_x = 0.
    \label{f2}
\end{equation}
The latter Eq is an eigenvalue equation for $B_x$.  With line tied boundaries, the eigenfunction and eigenvalue are
\begin{equation}
    B_x = \cos(Kx),\text{       }\gamma^2 = - [K^2 + k^2].
    \label{f3}
\end{equation}

From Eq (\ref{f1}), as above, $n$ has the same parity as $B_x$.   However, this parity violates the zero derivative boundary condition required for $n$. (This is a manifestation of the "mismatch" discussed earlier).    This violation indicates that solutions near the boundaries must be reconsidered: that is, there may be boundary layers as $x \rightarrow \pm1$. In what follows, we show that indeed there is a boundary layer.  A boundary layer in solution generally indicates rapid variation in the x-direction, i.e., $d/dx \gg 1$. As a consequence, there may be sharp variations in the eigenfunction, or, alternatively, a multiple scale character to the solution \citep{Bender}.  For this paper, we have investigated both possibilities and we find that the latter character, multiscale behaviour, is the apt description.  We proceed below with a multiscale analysis.

We go back to Eq (\ref{2n}). We note that $\gamma^2$ is already known to be $O(1)$.  But this time we order $\partial/\partial x$ large.  In that case, the term $T\partial_{x}^2$ must be taken to be optimally $O(1)$, and, so, retained in the equation. Thus, the multiple scale equation from Eq (\ref{2n}) remains
\begin{equation}
    (\gamma^2 - T\partial_x^2)n = \gamma^2 B_x. 
    \label{f4}
\end{equation}

Assuming even $n(x)$, the general solution to this inhomogeneous equation can be written
\begin{equation}
    n(x) = \cos(Kx)  -  A\cos(\kappa x), \text{         }  \kappa \gg K. 
    \label{f5}
\end{equation}
(Note that the particular solution is approximate:  a term of $O(K^2/\kappa^2)$ has been discarded.)  The constant $A$ is fixed upon demanding zero derivative of $n$ at $x=-1$.  Thus,
\begin{equation}
    n(x) = \cos(Kx)  -  [(K/\kappa)/\sin(\kappa)]\cos(\kappa x),        \text{         } \kappa \gg K.
    \label{f6}
\end{equation}

\begin{figure}
  \centering
  \includegraphics[width=0.6\textwidth]{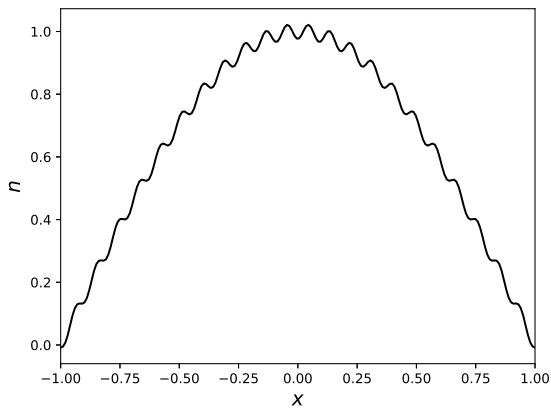}
  \caption{$n(x)$ plotted from Eq (\ref{f6}), at $T=0.2$ and $k=20$.  Sharp structured ripple riding on long wavelength envelope of $n(x)$ evident.  Note $dn/dx$ is forced to be zero at the $x=\pm1$ plates.}
\label{fig:fastmode}
\end{figure}
This solution is plotted in Fig.~\ref{fig:fastmode}, showing clearly the short wavelength ripple riding on the long wavelength solution for $n$.   Note also the zero derivative at the boundaries.  Note further that the ripple is of short wavelength, by a factor $\kappa/K$, but also that the amplitude of the ripple is small, by the same factor.  It can also be noted that the $A$cos$(\kappa x)$ part of the solution is sub-dominant in the region away from $x \rightarrow \pm1$, but that it becomes comparable to the particular solution, cos$(Kx)$, in the boundary layers.  This gives a narrow layer width, narrower by the same factor $(\kappa/K)$ as above.    The structure of the boundary layer can be seen upon expanding (\ref{f6}) near the boundaries.  At the left boundary, we let $x=-1+s$, where $s \ll 1$ and $\kappa s \sim 1$.  Equation (\ref{f6}) then becomes
\begin{equation}
    n(x \rightarrow -1) = K[s - (1/\kappa)\sin(\kappa s) -(1/\kappa)(\cos\kappa/\sin\kappa)].
    \label{f7}
\end{equation}
We conclude that the ripple is part of the eigenmode, and the “noise” must be concluded to be real.  In particular, in a numerical solution, it must be resolved.

\section{Numerical solutions}
\begin{figure}
  \centering
  \includegraphics[width=0.9\textwidth]{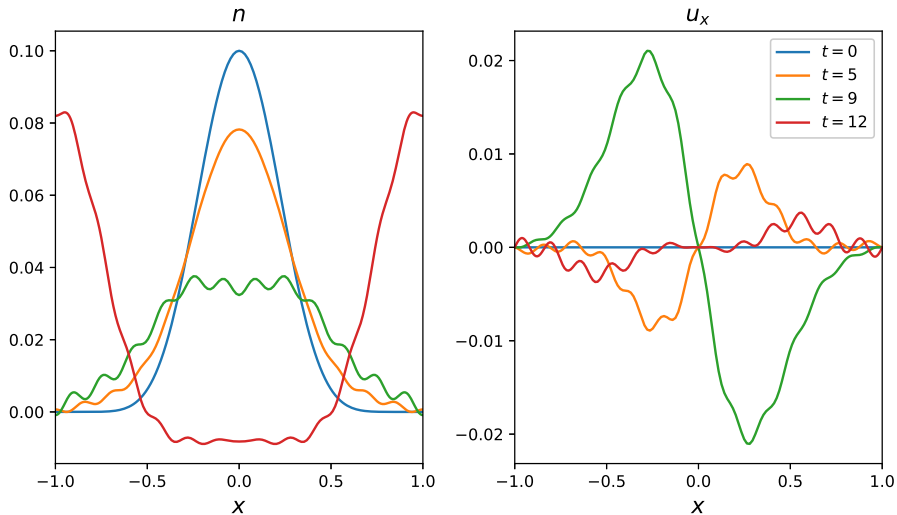}
  \caption{Solutions of $n$ and $u_x$ at $t=0,5,9,12$ from the  $N=200$ Chebyshev run.}
\label{fig:cheb}
\end{figure}

As a more in-depth examination of the noise, we conducted a series of numerical experiments with high resolution, using various boundary conditions in a finite-difference code, and increasing the number of grid points.  We also compared the finite-difference code with a much more accurate solution using a Chebyshev spectral method, with which the discretization error decreases exponentially when the number of collocation points $N$ increases \citep{Trefethen2000}.  We subjected the linear equations, (\ref{linear_n})-(\ref{linear_bx}), to an initial condition given by $n=((1+\cos(\pi x))/2)^4\exp(ikz)$. The temperature $T=0.2$ and the wavenumber $k=15$. Our primary objective was to quantify the numerical errors due to different implementations of spatial disretization; hence, we kept the timestep $\Delta t=5\times 10^{-5}$ for all calculations, with a trapezoidal leapfrog scheme for time stepping.    
Here we first report the solutions from a simulation using the Chebyshev code with $N=200$. By varying $N$ for a convergence test, we confirmed that the discretization error was reduced to approximately the level of floating point round-off error at this resolution. We show in Fig.~\ref{fig:cheb} traces of $n$ and $u_x$ at $t=0$, $5$, $9$, and $12$.  We note the early time traces are fairly smooth, and of long wavelength, but the late time solutions can be construed as “noisy”.  Of particular note is the highly “noisy” trace of $u_x$, and the short-scale “ripple” riding on top of the long wavelength envelope of the density profile.  

\begin{figure}
  \centering
  \includegraphics[width=1.0\textwidth]{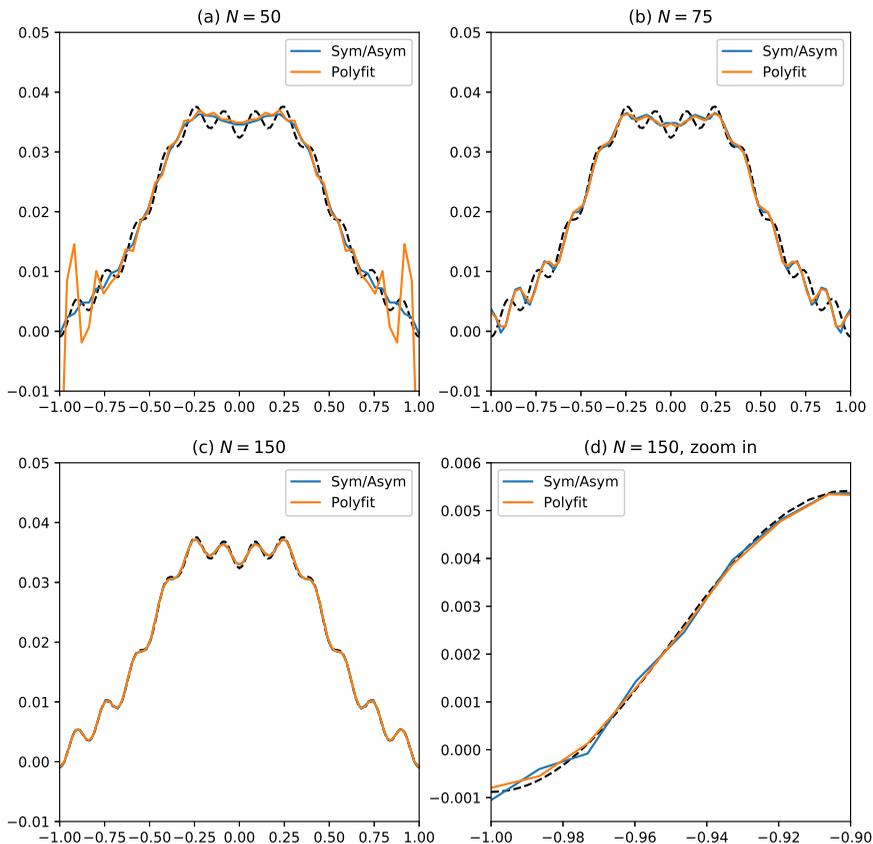}
  \caption{Finite-difference solutions at $t=9$ with various numbers of grid points $N$ and boundary conditions. Black dashed lines are the $N=200$ Chebyshev solution.}
\label{fig:FD-sol}
\end{figure}

Now we turn to the finite-difference implementations of the linear problem. Here we use a fourth order finite-difference scheme that requires two "ghost" points outside the boundary. In one set of calculations, we implemented the symmetric ($n$)/anti-symmetric ($u_x$, $u_z$, $B_x$) boundary conditions for the ghost points. However, we recall, our analysis of Secs. 4 and 5 above had revealed boundary layers, thus rapidly varying functions, bereft of any simple symmetries across the boundaries.   To address this behaviour, we tried also, in another set of calculations, to numerically fit the two ghost points to a solution using polynomials. For example, for the boundary condition $f(b)=0$, we used
\begin{equation}
   f = Ax + Bx^2. 
   \label{polyfit}
\end{equation}
This gave the following conditions for the ghost points, viz
\begin{equation}
    f_{-1}   =   - 3 f_1  +  f_2  
\end{equation}
\begin{equation}
    f_{-2}   =    - 8 f_1 + 3 f_2.
\end{equation}
Likewise, for $(df/dx)(b)=0$ boundary conditions, we fitted to
\begin{equation}
f = f_0 + Ax^2 + Bx^3.  
\end{equation}
This gave the following conditions for the ghost points, viz
\begin{equation}
f_{-1}   =   - (3/2) f_0  + 3 f_1 - (1/2) f_2
\end{equation}
\begin{equation}
f_{-2}   =   - 12 f_0  + 16 f_1 - 3 f_2.  
\end{equation}

We then re-ran our test example with the two sets of boundary conditions. Fig.~\ref{fig:FD-sol} shows the solutions at $t=9$ with various numbers of grid points $N$. Here, the black dashed line in each panel shows the Chebyshev solution with $N=200$, the blue line shows the solution with symmetric/anti-symmetric boundary conditions, and the orange line shows the solution with polynomial-fit boundary conditions. Panel (a) shows the cases with $N=50$. Here, the polynomial-fit solution exhibits high-amplitude fluctuations near the boundaries, because polynomial extrapolation can be very unreliable when the solution is under-resolved. Panel (b) shows the cases with $N=75$. Now the two sets of boundary conditions yield very similar solutions, and the solutions appear to be "noisy". Note that the fluctuations in the finite-difference solutions are out of phase with respect to the Chebyshev solution, due to the finite-difference discretization error. Panel (c) shows the cases with $N=150$, where the finite-difference solutions now approach the Chebyshev solution. While the two finite-difference solutions are rather similar, the close-up view near the boundary at $x=-1$, shown in panel (d), reveals a notable difference. Here we note that the solution with  symmetric/anti-symmetric boundary conditions exhibits  low-amplitude fluctuations at the grid scale, as a consequence of the "mismatch" discussed in Sec.~\ref{sec:mismatch}. In contrast, the solution with polynomial-fit boundary conditions is noticeably smoother. 

\begin{figure}
  \centering
  \includegraphics[width=0.6\textwidth]{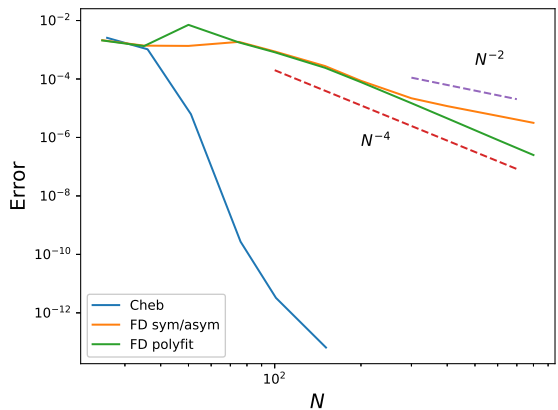}
  \caption{Scalings of errors with respect to $N$ for different implementations of spatial discretization.}
\label{fig:fig7}
\end{figure}

To quantify the error of finite-difference solutions, we measured the root-mean-square error of the density profile $n$ relative to the Chebyshev solution at $N=200$, which was taken as a proxy for the accurate solution. Fig.~\ref{fig:fig7} shows the scaling of errors as a function of $N$, for both sets of finite-difference solutions, as well as the errors of Chebyshev solutions. As a consequence of the exponential convergence of spectral methods, the error of Chebyshev solution decreases rapidly as $N$ increases. On the other hand, we expect the error of a fourth-order finite-difference method to decrease as $N^{-4}$. The $N^{-4}$ scaling is observed for both sets of boundary conditions at low $N$. However, the "mismatch" problem degrades the convergence rate to approximately $N^{-2}$ at high $N$ for the symmetric/anti-symmetric boundary conditions. In comparison, the $N^{-4}$ scaling is retained in the implementation of polynomial-fit boundary conditions. We conclude that the polynomial-fit boundary conditions are  superior to the symmetric/anti-symmetric boundary conditions, provided that the fluctuations of the solution are well-resolved.   



\section{Conclusion}
We have shown the appearance of rapidly varying functions in MHD situations which involve line-tying at boundaries.  Such short scale variations appear either as boundary layers, near the conducting boundaries, or as small amplitude ripples riding on the solution envelope.  The width of the boundary layers scales as the aspect ratio of the disturbance, ie, the ratio of layer width to parallel wavelength scales as  $k_{\parallel}/k_{\perp}$, where $k_{\parallel}$ and $k_{\perp}$ are parallel and perpendicular wavenumbers, and $k_{\parallel}/k_{\perp} \ll 1$. As far as the ripples, the ratio of ripple wavelength to parallel wavelength scales as $\sqrt{1/\beta}$.    Clearly, these smaller scales would have to  resolved. Nonlinear simulations we have performed, preliminarily, show that numerical instability can be triggered from the "noise" if there is insufficient resolution.  We note also that the ripple amplitude is small compared to the envelope, scaling as $\sqrt\beta$.

Our findings exemplify the caution of \cite{GreshoL1981}; to quote these authors:  "Don't suppress the wiggles -- they're telling you something!". Since rapidly varying solutions are a real consequence of line-tied boundary conditions, it is imperative that sufficient grid resolution be ensured and appropriate boundary conditions implemented in numerical calculations.  In particular,  diffusion layers near the boundary, if implemented, should be used with caution, depending on the importance of line-tying physics to the problem at hand.



\bibliographystyle{jpp}

\bibliography{linetied}

\end{document}